\documentstyle[prd,aps]{revtex} 
\begin{document}
\draft
\title{Diffusion of  the electromagnetic
energy due to the backscattering  off Schwarzschild
geometry.}

\author{Edward Malec}
\address{Physics Department, University College Cork, Cork,
Ireland and Institute of Physics, Jagellonian University, 
30-059 Krak\'ow, Reymonta 4, Poland}
\maketitle

\begin{abstract}
Electromagnetic waves propagate in the Schwarzschild spacetime like in
a nonuniform medium with a varying refraction index. A fraction of the
radiation scatters off the curvature of the  geometry. The energy of the
backscattered part   of an initially outgoing pulse of electromagnetic 
radiation can be  estimated, in the case of  dipole radiation, by a 
compact formula depending on the initial energy, the Schwarzschild radius
and  the pulse location. The magnitude of   the backscattered
energy depends on the frequency spectrum of the initial configuration.
This effect becomes negligible in the short wave limit,
but it can be significant in the long wave regime. Similar
results hold for the massless scalar fields and are expected to
hold also for weak gravitational waves.
\end{abstract}
\pacs{ }
\date{ }

\section{ Introduction.}

Backscattering  is a  phenomenon that prevents  waves from being
transmitted exclusively   along  null cones. That  aspect
of   waves propagation has been investigated for a long time
for various wave equations  (see, for instance,
\cite{Hadamard}). It  has been established that solutions of the Klein-Gordon
with nonuniform coefficients generically do exhibit backscattering
\cite{Hadamard}. This topic has been investigated in   general
relativity since early  1960's (\cite{MTW}, \cite{collective}
\cite{Bardeen1973}, \cite{Bonnor}, \cite{Gerhard}); a  comprehensive
bibliography can be found in \cite{Gerhard}.
The propagation of electromagnetic waves and of the resulting tails
has been studied in the early seventies (\cite{Bardeen1973} and
\cite{Rotenberg}) and recently by Ching et al. \cite{Ching} in the context
of the Schwarzschild spacetime and by Hod \cite{Hod}
in the context of the Kerr spacetime.
The backscattering effect can be understood as the result of  waves
propagation in  a nonuniform medium with a varying refraction index
\cite{refr}.

In \cite{MENOM}  has been assessed   a classical aspect
of the phenomenon that was not previously studied -
the energy diffusion through null cones - in the example
of a spherically symmetric  massless scalar field propagating in the
Schwarzschild geometry. The novel aspect of that work was
a  compact  estimate on the magnitude of the backscattered energy
in terms of the energy of initial data.

This paper is dedicated to the investigation of propagation of electromagnetic
fields in a background Schwarzschild spacetime. Similar to
\cite{MENOM}  the main attention is  focused on obtaining bounds on
the backscattered fraction of the  radiation energy, in terms of initial data.
>From the notional point of view the present paper parallels  \cite{MENOM},
with three notable exceptions. First, the crucial technical points of
the former work could have  been applied only to  spherically symmetric
fields. In order to overcome this difficulty, the  electromagnetic fields
have to be split, with the extraction of a known part which defines
initial data.  Then the standard expansion in terms of vector spherical
harmonics leads a problem that can be tackled with methods  applied
earlier in  \cite{MENOM}. Second,  an energy inequality is proven.
Third, this paper shows     that the  energy diffusion
 depends on the frequency of the radiation. An   example
of a  dipole radiation allows one to   characterize  this
quantitatively. The magnitude of the
backscattering can be characterized as the ratio of the backscattered
energy versus the  initial energy of outgoing waves.
This  is vanishingly small in the short wave regime but it can
be quite significant  in the long part of the radiation spectrum. This
kind of dependence on the    frequency can be expected to hold also
for higher multipoles. The scale is essentially set by the gravitational
radius of the gravity source. All results of this paper hold true for any
material sources of the Schwarzschild geometry - including stars, white
dwarves, neutron stars and black holes, although the effects can really
matter only in the two latter classes of objects.

The order of the remaining parts of this paper is following. The next section
defines  notation, basic equations  and a decomposition of  the
electromagnetic potential. The subsequent sections  of this
 work deal only  with   dipole
radiation. In section III is derived an energy estimate. Section IV is
dedicated to the derivation of a bound,
depending on the initial  energy, of the backscattered part of the potential.
Section V is devoted to the derivation of  useful estimates of a pair
of null-line integrals.
In Section VI the equations are formulated in the language of
characteristics.  Previously   found restrictions on the backscattered
part of the potential allow one to estimate
radiation intensities.  Section VII brings an improved estimate of the
backscattered potential, again basing on the method of characteristics.
The next Section proves    the main results - a bound on that
fraction of the energy that can diffuse due to the backscatter off
the Schwarzschild geometry curvature.    Section IX  shows that
in the case of short-wave radiation the   dipole radiation
backscatter is negligible. In contrast, in the
long-wave  regime the effect can be significant. Section X discusses
how the effect  depends on a distance and evaluates the exactness of
the obtained criteria.
The last Section presents a short summary and conclusions.

\section{Formalism}

Spherically symmetric geometry outside matter is given by
a Schwarzschildean geometry  line element,
\begin{equation} ds^2 = - (1-{2m\over R})dt^2 +
{1\over 1-{2m\over R}} dR^2 +
R^2 d\Omega^2~,
\label{1}
\end{equation}
where $t$ is a time coordinate, $R$ is a radial
coordinate that coincides with the areal radius
and $d\Omega^2 = d\theta^2 + \sin^2\theta d\phi^2$
is the line element on the unit sphere, $0\le \phi < 2\pi $
and $0\le \theta \le \pi $.

As it concerns the electromagnetic fields,
it is convenient to assume that the scalar component of the electromagnetic
potential
vanishes while the vector potential satisfies the Coulomb gauge condition.
Using a multipole expansion of the electromagnetic vector potential  in terms
of vector spherical harmonics one obtains \cite{Wheeler1955}
\begin{equation}
(-\partial_0^2 + \partial_{r^*}^2)\Psi_l = (1-{2m\over R}){l(l+1)\over R^2}
\Psi_l.
\label{2}
\end{equation}
$\Psi $'s should be essentially two-index functions, $\Psi_{lM}$,
(where $M$ is the projection of the angular momentum), but since
the evolution equation is $\phi -$ independent, the index $M$ is suppressed.
The variable $r^*\equiv R+2m\ln ({R\over 2m}-1)$ is the Regge-Wheeler tortoise
coordinate. The backreaction exerted  by the  electromagnetic field onto
the metric has been neglected in the present analysis. That is readily
 justified for any gravitational sources other than black holes.
In the case of a black hole  this  approximation holds true  some
 distance away from  its horizon \cite{MENOMback}.

Consider a set of functions  of the  form
\begin{equation}
\tilde \Psi_l(t, r^*)=\sum_{s=0}^l{\Psi_{ls}(r^* -t)\over R^s}
\label{3}
\end{equation}
where the functions $\Psi_{ls}$ are given by the recurrence relations
\begin{eqnarray}
&&\partial_{r^*}\Psi_{l1}=-{l(l+1)\over 2}\Psi_{l0}\nonumber\\
&& \partial_{r^*}\Psi_{l(s+1)}={
1\over 2(s+1)}\Bigl[ \bigl( s(s+1)-l(l+1)\bigr) \Psi_{ls}
-2m(s^2-1)\Psi_{l(s-1)}\Bigr] .
\label{4}
\end{eqnarray}
 In  \cite{MTW} is shown a dipole solution of this type \cite{div}.
In the Minkowski space-time (m=0) $\tilde \Psi_l$  solves (\ref{2});
it represents a purely outgoing electromagnetic radiation.

Let a function $\tilde \Psi_l$ be given by (\ref{3}) and (\ref{4}) and
assume   that (for space-like sections with $t\ge 0$) its  support  is
compact and located entirely in the vacuum region outside some radius
 $a>2m$, i. e., outside
the Schwarzschild radius. Let initial data
of a  solution $\Psi_l$ of (\ref{2}) concide with  $\tilde \Psi_l$ at $t=0$.
Thus initially $\Psi_l$ is a purely outgoing partial wave.
It should be noted that the assumption   that initial data are
(initially) purely  outgoing is made in this paper  only for the
sake of clear presentation. The propagation of
electromagnetic waves is a linear process as far as the backreaction can be
neglected. Therefore the propagation of the  initially outgoing radiation
(or even of a fraction of the outgoing radiation)
is independent of whether or not  the ingoing radiation is present.

It will be convenient to  decompose  the sought  solution $\Psi_l(r^*,t)$
into the known part $\tilde \Psi_l$ and an unknown function $\delta_l$
\begin{equation}
\Psi_l=\tilde \Psi_l +\delta_l.
\label{5}
\end{equation}
Initially $\delta_l=\partial_0\delta_l=0$.
A similar   splitting is done in \cite{Bardeen1973}, who
then seek  a series expansion  of $\delta_l $. This will be avoided
in the paper, in favour of finding a number of estimates of $\delta_l$ that
would provide the needed  information about the backscattered
part of the radiation.

In the rest of this paper will be considered only  dipole radiation,
$\Psi_1$.   Consequently, all angular momentum related
 subscripts will be omitted.

\section{An energy estimate}

The dipole term constitutes the most important part of the electromagnetic
radiation. Assume   dipole-type initial data
\begin{equation}
\tilde \Psi (x(R))=-\partial_{r^*} f(x(R))+{ f(x(R))\over R(r^*)},
\label{06}
\end{equation}
with the initial  support  $(a, \infty )$)
of a $C^2$-differentiable $ f$ and $x(R)
\equiv r^*(R)-r^*(a)$.
The differentiability  of $f$
guarantees that the initial  energy density  is continuous and vanishes on
the boundary $a$.

{\bf Lemma 1.} Define  $I_{a,\epsilon }(R)$
\begin{eqnarray}
 I_{a,\epsilon }(R)\equiv  \int_{a}^Rdr { f^2(x(r))\over r^{4+2\epsilon}},
\label{07}
\end{eqnarray}
and
\begin{eqnarray}
 \beta_a (R)\equiv  \int_{a}^Rdr { \tilde \Psi^2(x(r))\over r^2},
\label{08}
\end{eqnarray}
where $0<2\epsilon <1$.
Then for $a>2m(1+1/\sqrt{1+2\epsilon })$ the  following inequality holds
\begin{eqnarray}
I_{a,\epsilon }(R) \le {\beta_a (R)\over \epsilon a^{2\epsilon }}
{1-\Bigl({a\over R}\Bigr)^{2\epsilon }
\over (1+2\epsilon )(1-{2m\over a})^2-{4m^2\over a^2} }.
\label{08a}
\end{eqnarray}

{\bf Remark.}
The integral $\beta_a (R)$ is bounded above by the electromagnetic
energy $E_R(t)/(4\pi )$ defined later. Therefore
$$I_{a,\epsilon }(R) \le {E_a(R,t) \over 4\pi \epsilon a^{2\epsilon }}
{1-\Bigl({a\over R}\Bigr)^{2\epsilon }
\over (1+2\epsilon )(1-{2m\over a})^2-{4m^2\over a^2} }.$$

{\bf Proof.}

Notice that
\begin{eqnarray}
&&{ f^2\bigl( x(r)\bigr) \over r^2}
= \Biggl[ \int_a^rds \partial_s{f(s)\over s} \Biggr]^2
= \nonumber\\
&&\Biggl[ \int_{a}^rds {1\over 1-{2m\over s}}\Bigl( -{\tilde
\Psi \over s}+{2mf\over s^3}\Bigr) \Biggr]^2   \le
 \nonumber\\
&& 2\Biggl[ \int_{a}^rds {-\tilde\Psi \over s(1-{2m\over s})}\Biggr]^2
+2\Biggl[ \int_{a}^rds{2mf\over s^3(1-{2m\over s})}\Bigr) \Biggr]^2;
\label{09}
\end{eqnarray}
the second inequality follows from $(A+B)^2\le 2A^2+2B^2$. The factor
$1/(1-2m/s)$ that appears in the integrands
can be bounded above by $1/\eta_a$, where
\begin{equation}
\eta_a\equiv 1-{2m\over a}.
\label{010}
\end{equation}
Subsequently, the use of the
Schwarz inequality and simple integrations yield
\begin{eqnarray}
&&{ f^2\bigl( x(r)\bigr) \over r^2}
\le      \nonumber \\
&& {2\beta_a (r) (r-a)\over \eta_a^2}
+{8m^2I_{a,\epsilon }(r)\over \eta_a^2(1-2\epsilon )}
\Bigl( {1\over a^{1-2\epsilon }}-{1\over r^{1-2\epsilon }}\Bigr) .
\label{011}
\end{eqnarray}
The insertion of (\ref{011}) into the integral of (\ref{07}) gives
\begin{eqnarray}
&&I_{a,\epsilon }(R)
\le \nonumber\\
&& \int_{a}^Rdr{1\over r^{2+2\epsilon }}\Biggl[ {2\beta_a (r)
(r-a)\over \eta_a^2}
+{8m^2I_{a,\epsilon }(r)\over \eta_a^2(1-2\epsilon )}
\Bigl( {1\over a^{1-2\epsilon }}-{1\over r^{1-2\epsilon }}\Bigr)
 \Biggr] .
\label{012}
\end{eqnarray}
$\beta_a(r)$ and $I_{a,\epsilon }(r)$ are nondecreasing functions,
therefore taking them in front of the appropriate  integrals would not
make the corresponding terms  smaller. Straightforward integration of the
obtained expressions yields
\begin{eqnarray}
&&I_{a,\epsilon }(R)\le  {\beta_a (R)
\over \eta_a^2a^{2\epsilon }}\Biggl(
{1\over 2\epsilon }\bigl( 1-\Bigl({a\over R}\Bigr)^{2\epsilon } \bigr)
+ {1\over 1+ 2\epsilon }
\bigl(-1+\Bigl({a\over R}\Bigr)^{1+2\epsilon } \bigr) \Biggr)
 +\nonumber \\
&&
{2I_{a,\epsilon }(R)\over 1-2\epsilon }
 \Bigl( {2m\over a-2m}\Bigr)^2
\Biggl( {1\over  1+2\epsilon }
\bigl( 1-\Bigl({a\over R}\Bigr)^{1+2\epsilon } \bigr) -{1\over 2}
\bigl( 1-\Bigl({a\over R}\Bigr)^{2  } \bigr) \Biggr)
\label{013a}
\end{eqnarray}
One should note that  the expression inside the bracket in the first
line  can be estimated as follows
$${1\over 2\epsilon }\bigl( 1-\Bigl({a\over R}\Bigr)^{2\epsilon } \bigr)
+ {1\over 1+ 2\epsilon }
\bigl(-1+\Bigl({a\over R}\Bigr)^{1+2\epsilon } \bigr) \le
{1\over \epsilon (1+2\epsilon )}
\bigl( 1-\Bigl({a\over R}\Bigr)^{2\epsilon } \bigr) ,$$
while the  expression  inside the bracket  in the second line  is bounded
above by
$$ {1 -2\epsilon \over  2(1+2\epsilon) }
\bigl( 1-\Bigl({a\over R}\Bigr)^{1+2\epsilon } \bigr) . $$
Eq. (\ref{013a}) can be now written as
\begin{eqnarray}
I_{a,\epsilon }(R)\le {\beta_a (R)
\bigl( 1-\Bigl({a\over R}\Bigr)^{2\epsilon } \bigr)
\over \eta_a^2a^{2\epsilon }\epsilon (1+2\epsilon)}
+\Bigl( {2m\over a-2m}\Bigr)^2{I_{a,\epsilon }(r)\over 1+2\epsilon }.
\label{013}
\end{eqnarray}
Rearranging Eq. (\ref{013}) so that the two terms with $I_{a,\epsilon }(R)$ are
on the left hand side, one obtains
\begin{equation}
I_{a,\epsilon }(R)\times \Biggl( 1- \Bigl({2m\over a-2m}\Bigr)^2 {1\over 1+2\epsilon}\Biggr)
\le {\beta_a (R)
\bigl( 1-\Bigl({a\over R}\Bigr)^{2\epsilon } \bigr)\over \eta_a^2a^{2\epsilon }\epsilon (1+2\epsilon)} ;
\label{014}
\end{equation}
this gives the postulated  bound  of $I_{a,\epsilon }(R)$  if
$a>2m(1+1/\sqrt{1+2\epsilon })$, as assumed above.

The obtained formula is not exact, but  with the appropriate choice of
 $f$ and $\epsilon $ the error is small. Take for instance
$f=C $ within $(a+a_1, b-b_1)$, $a_1, b_1<<a$, $b>>a$ (which obviously
means that  $b-a>>a_1, b_1$), and let
$f$ be smoothly joined  to zero outside $(a,b)$ by some intermediary
functions. Under those conditions, a direct calculation gives
\begin{equation}
{I_{a,\epsilon }(b)\over \beta_a(b) }\approx {3\over (3+2\epsilon)a^{2\epsilon }};
\end{equation}
as compared with ${1\over \epsilon (1+\epsilon )a^{2\epsilon }}$ that
follows from  (\ref{014}). If $\epsilon \approx 1/2$, then the exact
result  differs by less than 25 percent from the bound of (\ref{014}).
Later on will be used the value $\epsilon =1/8$ (which appears to be
more economical in subsequent calculations), in which case  the above
estimate deteriorates significantly. The exact value of
${I_{a,\epsilon }(b)\over \beta_a(b) }$  is then roughly 15 \% of that predicted by
 (\ref{014}).

\section{Estimating $\delta $.}

$\delta$ is initially  zero and its evolution is governed by the following
equation
\begin{equation}
(-\partial_0^2 + \partial_{r^*}^2)\delta = (1-{2m\over R})
\Biggl[ { 2\over R^2}
\delta + {6mf\over R^{4}}  \Biggr] .
\label{6}
\end{equation}
Define $\tilde \Gamma_{(R,t)} $ - a null geodesic that originates
at $(R,t)$ and is directed outward. If a  point lies on the initial
hypersurface, then $\tilde \Gamma_{(R,0)}\equiv  \tilde \Gamma_R$.
By $\tilde \Gamma_{(R_0, t_0), (R,t)}$ will be understood a segment of
 $\tilde \Gamma_{(R_0, t_0)}$ ending at $(R,t)$.

Later  will be needed   a following bound.

{\bf Theorem 2.}   Let  the support of initial data  be $(a, b)$,
$b\le \infty $ and let   $\tilde \Gamma_{R_0, (R,t)}$  be   the outgoing null
geodesic from    $(R_0,t=0)$  to $(R,t)$. Then
\begin{equation}
{|\delta(R)|\over R}\le
m C_1 \sqrt{\beta_a(b)} {1\over a^{\epsilon }\sqrt{R\eta_{_R}}}
\bigl(  {1\over R_0^{1-\epsilon }}-{1\over R^{1-\epsilon }}\bigr) ,
\label{7}
\end{equation}
where
\begin{equation}
 C_1 \equiv
{6\sqrt{2} \over\eta_a^{3/2}(1-\epsilon)}
\sqrt{{1\over \epsilon \bigl[ (1+2\epsilon )\eta_a^2-{4m^2\over a^2}
\bigr] }}
\label{7a}
\end{equation}
and $\eta_{_R}=1-2m/R$.

{\bf Proof.}
Define an energy  $H(R,t)$ of the  field $\delta$
contained in the exterior of a sphere of a radius $R$,
\begin{equation}
H(R,t) = \int_{R}^{\infty }dr
\Bigl(  {(\partial_0\delta )^2\over 1-{2m\over r}} + (1-{2m\over r})
(\partial_r\delta)^2+(\delta)^2{ 2\over r^2}\Bigr) .
\label {8}
\end{equation}
One can  easily show that
\begin{eqnarray}
&&
(\partial_t+\partial_r^*)H(R,t)= \nonumber\\ &&
-(1-{2m\over R})\Biggl[ (1-{2m\over R})
\Bigl( {\partial_0\delta \over (1-{2m\over R})} +
\partial_R\delta
 \Bigr)^2 +{ 2\over R^2}\delta^2 \Biggr] -
12m \int_R^{\infty }dr\partial_0\delta
{ f\over r^{4}} \le
\nonumber\\ &&
-12m \int_R^{\infty }dr\partial_0\delta
{ f\over r^{4}}
\label{9}
\end{eqnarray}
the inequality follows from omission of the nonpositive boundary term.
The right hand side can be bounded further by
\begin{equation}
12m\Bigl[ \int_R^{\infty }dr(\partial_0\delta)^2\Bigr]^{1/2}
\Bigl[ \int_R^{\infty }dr { f^2\over r^8} \Bigr]^{1/2},
\label{10}
\end{equation}
due to the Schwarz inequality. That in turn can be bounded by
\begin{equation}
{12m\over R^{2-\epsilon }\sqrt{\eta_R }}H^{1/2}
\Bigl[ \int_R^{\infty }dr { f^2\over r^{4+2\epsilon }}  \Bigr]^{1/2}.
\label{11}
\end{equation}
The second integral in (\ref{11}) can not increase along outgoing null
directions, therefore is bounded by initial values,
$ \int_R^{\infty }dr  { f^2\over r^{4+2\epsilon }} \le
\int_{R_0}^{\infty }dr  { f^2\over r^{4+2\epsilon }}\equiv  I_{R_0,
\epsilon }(R)$. Since $I_{R_0, \epsilon }(\infty )\le
I_{a,\epsilon }(\infty )$, one arrives at
\begin{equation}
(\partial_t+\partial_r^*)H(R,t)^{1/2}\le  6\sqrt{I_{a,\epsilon }(\infty )}
{m\over \sqrt{\eta_a }R^{2-\epsilon }}.
\label{12}
\end{equation}
The integration of (\ref{12}) along $\tilde \Gamma_{R_0, (R,t)}$ yields,
replacing $I_{a,\epsilon }(\infty )$ by its bound expressed in (\ref{08a}),
\begin{eqnarray}
&&\sqrt{H(R,t)}  \le   {6m\over a^{\epsilon }}
{ \sqrt{\beta_a(\infty )}\over \eta_a^{3/2}(1-\epsilon)}
\sqrt{{1\over \epsilon
\bigl( ( 1+2\epsilon )\eta_a^2-{4m^2\over a^2}\bigr) }}
\bigl(  {1\over R_0^{1-\epsilon }}-{1\over R^{1-\epsilon }}\bigr) =
\nonumber\\
&& ={mC_1\over \sqrt{2} a^{\epsilon } }
\bigl(  {1\over R_0^{1-\epsilon }}-{1\over R^{1-\epsilon }}\bigr) .
\label{13}
\end{eqnarray}
Notice that initially $\delta$ vanishes and that its
propagation is ruled by a  hyperbolic equation.  Thence
at any finite time $t$ the  support of $\delta $ is bounded. Therefore
\begin{eqnarray}
&&{|\delta(R)|\over R}=|\int_{\infty}^R\partial_r{\delta(r)\over r}|\le
\nonumber\\
&& \bigl( \int_{\infty}^R{1\over r^2}\bigr)^{1/2}
\bigl( \int_{\infty}^R(\partial_r\delta-{\delta\over r})^2 \bigr)^{1/2}
\nonumber\\
&&\le   {1\over \sqrt{R-2m}}(2H)^{1/2}(R)
\label{14}
\end{eqnarray}
Inequalities (\ref{13}) and  (\ref{14}) yield the bound of Theorem 2
in the case when $b=\infty $.

Let the initial data be of  finite support $(a, b)$. Define a region
$\Omega_b$ consisting of points $(R\ge b, t)$ acausal to $(b, t=0)$.
The energy $H(b(t), t)$ obviously vanishes for any point $(b(t), t)$
located  inside  $\Omega_b$. In this case the inequality of
Theorem 2 can be stated as follows:
\begin{equation}
{|\delta(R)|\over R}\le
mC_1 \sqrt{\beta_a(b)} {\sqrt{1-({a\over b})^{2\epsilon }}\over
 a^{\epsilon }\sqrt{R
\eta_R}}
\bigl(  {1\over R_0^{1-\epsilon }}-{1\over R^{1-\epsilon }}\bigr) ,
\label{7b}
\end{equation}
In what follows it will be assumed that the initial data have   compact
support located in an annular region $(a, b)$.

\section{Estimates of  two (null) line integrals}

In analogy with $\tilde \Gamma_{(R,t)}$ defined earlier, let
$ \Gamma_{(R,t)}$  be  a null ingoing geodesic that originates at $(R,t)$.
$\Gamma_{(R,t=0)}$ will be shortened to $ \Gamma_{R}$. A segment of
$ \Gamma_{(R_1,t_1)}$ connecting $(R_1,t_1)$ with $(R, t)$ ($t_1<t,
R_1>R$) will be denoted as $  \Gamma_{(R_1,t_1), (R, t)}$.

Let a  point  $(R, t)$ be an intersection of an ingoing null geodesic
$\Gamma_{R_1}$ with an outgoing null geodesic $\tilde \Gamma_a$.
Let $(r,\tau )$, $r \ge R$ be a point of
$\Gamma_{R_1,(R,t)}$  and define $(R_0(r), t=0)$ as a point
of the initial hypersurface such that
$\tilde \Gamma_{R_0}  \cap \Gamma_{R_1} =(r,\tau ) $.
Fixing $a$ and $R_1$, one can view $R_0$ as a function of $r$; obviously
$R_0(R)=a$ while $R_0(R_1)=R_1$.  On the other hand, fixing only $a$ and
viewing $R_1$ as a function of $R$ one has $R_1(a)=a$; this will be used in
the forthcoming proof.

One  can prove

{\bf Lemma 3}. Under above conditions and if  $R_1\le b$,
the line integral along a null
segment geodesic  $\Gamma_{R_1, (R, t)}$  is bounded above,
\begin{equation}
\int_R^{R_1} dr {r-R_0\over R_0r\sqrt{1-{2m\over R}}}\le
{1\over 2} \Bigl( \ln {R\over b}+\ln ({b-  2m \over a-2m}) \Bigr) .
\label{015}
\end{equation}
{\bf Lemma 4}. Under the above condition but with
 the initial point $(R_1,s)$ ($s>0$) of the null geodesic
segment  $\Gamma_{(R_1, s),(R,t)}$ lying   on
$\tilde \Gamma_{b}$ (Fig. 2), one can prove
\begin{equation}
\int_R^{R_1} dr {r-R_0\over R_0r\sqrt{1-{2m\over R}}}\le
{1\over 2}  \ln ({b- 2m \over  a- 2m}) \Bigr) .
\label{016}
\end{equation}
{\bf Proof of Lemma 3.}

Let  $r$ be a radial coordinate of a point  lying on the intersection of
$\tilde \Gamma_{R_0}$ and $\Gamma_{R_1}$ (Fig. 1). One finds that
the areal distances of three points $(R_0(r),0)$, $(r, \tau )$ and
$(R_1, 0)$ satisfy following
\begin{equation}
R_0(r)=2r-R_1+ 2m\ln \Bigl( {(r-2m)^2\over (R_0(r)-2m)(R_1-2m)}
\Bigr) .
\label{18c}
\end{equation}
That implies
\begin{equation}
dR_0=  2{1-{2m\over R_0}\over
1-{2m\over r}}dr.
\label{19}
\end{equation}
Replacing $r$ by $R_0$ in the integral of (\ref{015}) one obtains
\begin{eqnarray}
&&\int_R^{R_1} dr {r-R_0\over R_0r\sqrt{1-{2m\over r}}}  =
\int_R^{R_1} {dr  \over  \sqrt{1-{2m\over r}}}\Bigl( {1\over R_0}-
{1\over r}\Bigr) =
\nonumber \\
&&{1\over 2}\int_{a}^{R_1} dR_0{\sqrt{1-{2m\over r}}
\over R_0-2m}  -\int_R^{R_1} {dr \over r\sqrt{1-{2m\over r}}}\le
\nonumber \\
&&{1\over 2}\ln ({R_1-2m\over a-2m}) -\ln{R_1\over R}=\nonumber \\
&&\ln {R\over \sqrt{aR_1}} + {1\over 2}\ln {1-{2m\over R_1}\over 1-{2m\over a}}.
\label{20}
\end{eqnarray}
Next, one can show that $R_1\ge 2R-a$. Indeed, assuming that $a$
is fixed, one has from (\ref{19}) ${dR_1\over dR}\ge 2$; since
the initial condition is  $R_1(a)=a$, the conclusion   follows.

Taking into account  $R_1\ge 2R-a$  one  gets
\begin{equation}
\ln {R\over \sqrt{aR_1}} \le \ln {R\over \sqrt{a(2R-a)}}  \le
\ln \sqrt{R\over a }.
\label{017}
\end{equation}
Replacing $R_1$ by $b$ in the last term of (\ref{20})  and inserting
(\ref{017}), one arrives at the first of conjectured  inequalities,
(\ref{015}).

In order to prove  Lemma 4 one should    start  from
relation between areal distances of four points $(r, \tau)$,
$(R_0(r), 0)$, $(R,t)$ and $(a,0)$ (see
Fig. 1):
\begin{equation}
2\Bigl( r-R+2m \ln {r-2m\over R-2m}\Bigr) = R_0-a+2m \ln {R_0-2m\over a-2m};
\label{018}
\end{equation}
The variable $r$ ranges from $R_1>b$ to $R$.
Fixing  $a$ and $R$, one again  obtains
\begin{equation}
dR_0=  2{1-{2m\over R_0}\over
1-{2m\over r}}dr.
\label{020}
\end{equation}
A straightforward calulation, in which $dr$ is replaced by $dR_0$,
shows that
\begin{equation}
\int_R^{R_1} dr {r-R_0\over R_0r\sqrt{1-{2m\over R}}}\le
{1\over 2}  \ln ({b- 2m \over  a- 2m}) -\ln {R_1\over R} \Bigr) .
\label{021}
\end{equation}
Since $R_1 \ge R$, one immediately obtains (\ref{016}).

\section{An estimate of the    amplitudes   backscattered inward}

Define the intensity of the backscattered   radiation that is directed inward
\begin{equation}
h_-(R,t) ={1\over 1-{2m \over R}}(\partial_0+\partial_{r^*})\delta.
\label{15}
\end{equation}
Eq. (\ref{6}) reads now
\begin{equation}
(-\partial_0 + \partial_{r^*})\Bigl( (1-{2m\over R})h_-\Bigr)
= (1-{2m\over R})
\Biggl[ { 2\over R^2}
\delta + {6mf\over R^4} \Biggr] .
\label{16}
\end{equation}
The integral form of (\ref{16}) reads,
\begin{equation}
 (1-{2m\over R})h_-(R,t)
= \int_{R_1}^Rdr
\Biggl[ { 2\over r^2}
\delta + {6mf\over r^4} \Biggr] +h_-(R_1,s);
\label{17}
\end{equation}
here the integration  contour coincides with a null ingoing geodesics
$\Gamma_{(R_1,s),(R,t)}$. $(R_1,s)$ lies on the initial hypersurface
($s=0$) if $R_1\le b$;
thus   $h_-(R_1,s=0)=0$, since the initial data  are entirely outgoing.
If $R_1>b$ then $(R_1,s) \in \tilde \Gamma_{b, (R_1,s)}$; also
in this case $h_-(R_1,s)=0$, because $\tilde \Gamma_{b, R_1}$ constitutes the
outer boundary of the outgoing impulse. In either case the radiation
amplitude satisfies the integral equation
\begin{equation}
 (1-{2m\over R})h_-(R,t)
= \int_{R_1}^Rdr
\Biggl[ { 2\over r^2}
\delta + {6mf\over r^4} \Biggr] .
\label{17aa}
\end{equation}
The second term is bounded above by
\begin{eqnarray}
&&\int_{R}^{R_1}dr {6m|f|\over r^4} \le \nonumber\\ &&
6m\Biggl( \int_{R}^{R_1}dr  {f^2\over r^{4+2\epsilon }}\Biggr)^{1/2}
\Biggl( \int_{R}^{R_1}dr  {1\over r^{4-2\epsilon }}\Biggr)^{1/2}=
\nonumber\\
&& {6 m\over \sqrt{3-2\epsilon }}
\Biggl( \int_{R}^{R_1}dr  {f^2\over r^{4+2\epsilon }}\Biggr)^{1/2}
{1\over  R^{(3/2)-\epsilon }}\sqrt{1-{R^{3-2\epsilon }\over
R_1^{3-2\epsilon }}}\le
\nonumber\\
&& {6 m\over \sqrt{3-2\epsilon }}
\Biggl( \int_{R}^{R_1}dr  {f^2\over r^{4+2\epsilon }}\Biggr)^{1/2}
{1\over  R^{(3/2)-\epsilon }}\sqrt{1-{a^{3-2\epsilon }\over
b^{3-2\epsilon }}}
\label{17a}
\end{eqnarray}
where the second line follows from the Schwartz inequality and the last
inequality is due to the fact that $R/R_1\ge a/b$ (Appendix A).

In order to find the integral from the last line of
 of (\ref{17a}) it is useful to
project it   onto the initial data surface,
along outgoing null geodesics $\tilde \Gamma_{ R_0, (r, \tau)}$.
Notice that $dR_0=  2{1-{2m\over R_0(r)}\over (1-{2m\over r})}dr$ -
see  (\ref{19}). The $f^2/r^{4+2\epsilon }$
term cannot decrease during
this projection. One arrives at
\begin{equation}
\int_{R}^{R_1}dr  {f^2(r)\over r^{4+2\epsilon }} \le
\int_{a}^{R_1}dR_0{1-{2m\over r}\over
2(1-{2m\over R_0(r)})}{f^2(R_0)\over R_0^{4+2\epsilon }}
\le {I_{a,\epsilon }(R_1)\over 2 \eta_a}.
\label{17b}
\end{equation}
Inserting  the energy  estimate of Lemma 1 into (\ref{17b}) one  gets
finally
\begin{equation}
\int_{R}^{R_1}dr {6m|f|\over r^4}
   \le  \sqrt{\beta_a(b)}{mC_2\over  a^{\epsilon } R^{3/2-\epsilon }}
 \sqrt{1-({a\over b})^{2\epsilon }}.
 \label{022}
 \end{equation}
Here the constant $C_2$ is given by
\begin{equation}
C_2={3\sqrt{2\bigl( 1-{a^{3-2\epsilon }\over b^{3-2\epsilon }}
\bigr)}\over \eta_a^{3/2} \sqrt{\epsilon [(1+2\epsilon )
\eta_a^2-{4m^2\over a^2}](3-2\epsilon )}}.
\label{023}
\end{equation}
The $\delta $-related term of (\ref{17aa}) is bounded, due to (\ref{7}),
by
\begin{equation}
{2mC_1\over  a^{\epsilon } }\sqrt{\beta_a(b)}
 \sqrt{1-({a\over b})^{2\epsilon }}
\int_R^{R_1}dr{1\over \eta_Rr^{3/2}}\Bigl( {1\over R_0^{1-\epsilon }}
-{1\over r^{1-\epsilon }} \Bigr)
\label{024}
\end{equation}
Here $r\ge R_0$ and $r, R_0\in \tilde \Gamma_{R_0, (r,\tau )}$. Thus
$1/(r^{\epsilon } R_0^{1-\epsilon })\le 1/R_0$.
Therefore the  expression of (\ref{024}) is bounded above by
\begin{equation}
{2mC_1\over  a^{\epsilon }  }\sqrt{\beta_a(b)}
 {\sqrt{1-({a\over b})^{2\epsilon }}\over
R^{3/2-\epsilon } }
\int_R^{R_1}dr{1\over \eta_R}\Bigl( {1\over R_0}
-{1\over r} \Bigr) .
\label{025}
\end{equation}
Results of Lemma 3 and 4 lead now to a pair of estimates.
If $R_1\le b$ then
\begin{equation}
2\int_R^{R_1}dr{|\delta |\over r^2}\le
mC_1\sqrt{\beta_a(b)} {\sqrt{1-({a\over b})^{2\epsilon }}\over   a^{\epsilon }
R^{3/2-\epsilon } }\Bigl( \ln{R\over a} +
\ln {\eta_b\over \eta_a}  \Bigr)
\label{026}
\end{equation}
and if $R_1>b$ (in which case  $(R, t)\in   \Gamma_{  (R_1,s)}$) then
\begin{equation}
2\int_R^{R_1}dr{|\delta |\over r^2}\le
mC_1\sqrt{\beta_a(b)} {\sqrt{1-({a\over b})^{2\epsilon }}\over  a^{\epsilon }
R^{3/2-\epsilon } } \Bigl( \ln {b\over a} +\ln {\eta_b\over \eta_a}
 \Bigr) .
\label{027}
\end{equation}
In summary, the radiation amplitude is bounded above by
\begin{eqnarray}
&& (1-{2m\over R})|h_-(R,t)|  \le   \nonumber\\
&&{C_3\over  a^{\epsilon }R^{3/2-\epsilon } } \Biggl[ C_4+C_1
\ln {b\Theta (R_1(R)-b)+ R\Theta (-R_1(R)+b)  \over a} \Biggr] ,
\label{21}
\end{eqnarray}
where   $ \Theta (b-R_1) =0$ if $b-R_1 <0$ and  $ \Theta (b-R_1) =1$
if $  b-R_1 \ge 0$.
The constants $C_3$ and $C_4$ are  defined  by
\begin{eqnarray}
&& C_3\equiv
m\sqrt{\beta_a(b)} \sqrt{1-({a\over b})^{2\epsilon }},\nonumber\\
&&C_4\equiv C_2 +C_1\ln  {\eta_b\over \eta_a}.
\label{21a}
\end{eqnarray}

\section{Refining the bound on $\delta$}

Equation (\ref{15}) can be written in the integral form
\begin{equation}
\delta(R,t) =\int_{R_0}^{(R,t)}dr h_- +\delta (R_0),
\label{22}
\end{equation}
where the integration contour coincides with $\tilde \Gamma_{(R_0), (R,t) }$
and $R_0$  is a point of the initial Cauchy slice defined earlier.  Since
initially $\delta $ vanishes, one has  $\delta(R,t) =\int_{R_0}^Rdr h_-  $.
It becomes clear in Sec. VIII that  one needs to bound $\delta(R,t)$
only along  $\tilde \Gamma_{a,  \infty  }$; in what follows is always
meant this situation.
Define $R(b)\equiv (R: R_1(R)=b)$ (see Fig. 2).
Inserting the bound of (\ref{21}) (but notice that  (\ref{21})
bounds $\eta_r|h(r)|$, not $|h(r)|$ itself), one obtains
\begin{eqnarray}
&&|\delta(R)| \le \nonumber \\
&&{C_3 \over \eta_aa^{\epsilon }}  \Biggl[
C_4\int_a^{R }{dr\over r^{3/2-\epsilon } } +
 C_1\Theta (R(b)-R)\int_a^{R } dr{\ln {r\over a}\over  r^{3/2-\epsilon } } +
\nonumber \\
&&   C_1\Theta (R-R(b))\Bigl(
\int_a^{R(b)} dr{\ln {r\over a}\over  r^{3/2-\epsilon } } +\ln {b\over a}
\int_{R(b)}^{R }{dr\over r^{3/2-\epsilon } } \Bigr) \Biggr] .
\label{029}
\end{eqnarray}
The integrand of the second integral  is nonnegative, therefore extending
the integration up to $R(b)$ can give only a bigger quantity.
Thus one gets, after elementary integration,
\begin{eqnarray}
&&|\delta(R)| \le \nonumber \\
&&{2C_3 \over \eta_a(1-2\epsilon )a^{1/2}}  \Biggl[
C_4\Bigl( 1- \Bigl( {a\over R}\Bigr)^{1/2-\epsilon } \Bigr) +
+{2C_1\over 1-2\epsilon } \Bigl( 1- \Bigl( {a\over R(b) }
\Bigr)^{1/2-\epsilon }\Bigr) +\nonumber \\
&& C_1\Bigl( {a\over R(b) } \Bigr)^{1/2-\epsilon }
\Bigl( -\ln {R(b)\over a} +\Theta (R-R(b))\ln {b\over a}
\Bigl( 1- \Bigl( {R(b)\over R}\Bigr)^{1/2-\epsilon } \Bigr) \Bigr) \Biggr]
\label{030}
\end{eqnarray}
Dropping out the negative term  $-\Theta (R-R(b))  \ln {b\over a}
\Bigl( {R(b)\over R}\Bigr)^{1/2-\epsilon } $
and  taking into account  that $ -\ln {R(b)\over a} +
\Theta (R-R(b))\ln {b\over a}
\le \ln {b\over R(b)}$, one arrives  at
\begin{eqnarray}
&&|\delta(R)| \le \nonumber \\
&&{2C_3 \over \eta_a(1-2\epsilon )a^{1/2}}  \Biggl[
C_4\Bigl( 1- \Bigl( {a\over R}\Bigr)^{1/2-\epsilon } \Bigr)
+ \nonumber \\
&&{2C_1\over 1-2\epsilon } \Bigl( 1- \Bigl( {a\over R(b) }
\Bigr)^{1/2-\epsilon }\Bigr) +
C_1\Bigl( {a\over R(b) } \Bigr)^{1/2-\epsilon }
\ln {b\over R(b)}  \Biggr]
\label{031}
\end{eqnarray}
Define
\begin{equation}
\kappa \equiv (b-a)/a.
\label{032a}
\end{equation}
 One can show (see Appendix B) that
\begin{equation}
{a+b\over 2}- m\kappa  \le  R(b) \le   {a+b\over 2};
\label{032}
\end{equation}
the equality is achieved in the Minkowski space-time (m=0).
Since $b=a+a\kappa $, one has $R(b)\ge a+\eta_a\kappa /2$ or, defining
\begin{equation}
\alpha \equiv {\eta_a\over 2},
\end{equation}
$R(b)\ge a+\alpha  \kappa $.
The  insertion of the above into (\ref{031}) yields
\begin{eqnarray}
&&|\delta(R)|\le
{C_3 \over \eta_a(1-2\epsilon )a^{1/2}}\Biggl[ C_4
\Bigl( 1- \Bigl( {a\over R}\Bigr)^{1/2-\epsilon } \Bigr) +C_5\Biggr] ,
\label{033}
\end{eqnarray}
where
\begin{equation}
C_5\equiv   C_1{\ln {1+ \kappa \over 1+\alpha  \kappa }
\over \Bigl( 1+\alpha  \kappa \Bigr)^{1/2-
\epsilon }}+{2C_1\over 1-2\epsilon }\Bigl(1-
{1\over (1+  \kappa /2)^{1/2-\epsilon}} \Bigr) .
\label{034}
\end{equation}
This estimate   gives a better control
over the asymptotic behaviour of $\delta$  than  the former one,
(\ref{14}),  by a factor $1/\sqrt{R}$. In particular,
 now $\delta^2/R^2$ is known to be integrable. This integrability will
be exploited in the next section.

\section{Bounding the radiation energy loss }

The  energy  $E_R(t)$ of the electromagnetic field  $\Psi $
contained in the exterior of a sphere of a radius $R$ reads
\begin{equation}
E_R(t) =2\pi  \int_{R}^{\infty }dr
\Biggl(  {(\partial_0\Psi )^2\over 1-{2m\over r}} + (1-{2m\over r})
(\partial_r\Psi )^2+{2(\Psi )^2\over r^2}\Biggr) .
\label{035}
\end{equation}
Let  the initial data be as  specified hitherto, $\Psi (t=0)=\tilde \Psi $
and $\partial_0\Psi (t=0)=\partial_0\tilde \Psi $ for some $\tilde \Psi $;
thus  they  vanish outside an annular region
 $(a, b)$. Define $E_a^b=E_a(0)$ as the energy of the initial pulse.
 If  initial configuration is  purely outgoing, then
 $\partial_t\Psi = -\partial_{r^*}\tilde \Psi +f\partial_{r^*}(1/R)$.

Let  an outgoing  null cone $C_{a}$ originate from $(a,0)$.
In the Minkowski spacetime the outgoing radiation contained
outside $C_{a}$ does not leak inward and its  energy remains constant.
In a curved spacetime, however, some  energy
 will be lost from the main stream
due to the  diffusion
of the  radiation $h_-$ through   $C_{a}$.  Most of the
backscattered radiation will fall onto the center of the gravitational
attraction.  The forthcoming theorem gives a bound on the amount
 of diffused energy.

{\bf Theorem 3.}  Under the above assumptions,
 the fraction of the diffused energy  $\delta E_a/E_a^b $
satisfies the inequality
\begin{eqnarray}
&&{\delta E_a\over E_a^b} \le  \Bigl( {2m\over a}\Bigr)^2
{1-{1\over (1+ \kappa  )^{2\epsilon }}\over  \eta^2_a  }\times
 \nonumber\\ &&
 \Biggl[ { C_4^2\over   (1-\epsilon)}\Bigl( {\eta_a  \over 16} +
 {2\epsilon^2\over  (1-2\epsilon )^2
 (3-2\epsilon )}\Biggr) +{C^2_5
 \over (1-2\epsilon )^2}+C_6+
{ 2C_4C_5  \over (1-2\epsilon )(3-2 \epsilon )} \Biggl] ,
\label{036}
\end{eqnarray}
where  $C_1$ - $C_5$ have been defined earlier and
\begin{eqnarray}
&&C_6={\eta_a C^2_1\over 16(1-\epsilon )}
\Bigl( {\ln^2{1+ \kappa \over 1+\alpha  \kappa }\over (1+\alpha
\kappa )^{2-2\epsilon }}
+{1-{1\over (1+ \kappa /2)^{2-2\epsilon }}\over (1-\epsilon )^2}\Bigr) +
\nonumber \\
&& {C_1C_4\eta_a \over 16 (1-\epsilon )}  \Bigl(
{(2+{C_1 \ln (1+ \kappa /2)
\over C_4(1-\epsilon )})\ln {1+ \kappa \over 1+\alpha  \kappa }\over
(1+\alpha  \kappa )^{2-2\epsilon }}+
{1-{1\over (1+ \kappa /2)^{2-2\epsilon }}\over (1-\epsilon )}\Bigr) .
\label{037}
\end{eqnarray}
{\bf  Proof. }
 The rate  of the energy change along $C_{a}$ is given by
\begin{eqnarray}
&&(\partial_0+\partial_{r^*})E_{{a}}= \nonumber\\
&& -2\pi (1-{2m\over R})\Biggl[ (1-{2m\over R})\Bigl( {\partial_0\Psi \over
1-{2m\over R}}
 +\partial_R\Psi \Bigr)^2  +{ 2\over R^2}\Psi^2 \Biggl] =
\nonumber\\ &&
-2\pi (1-{2m\over R})\Biggl[ (1-{2m\over R}) \Biggl( h_--
 { f\over R^{2}}\Biggr)^2
  +{ 2\over R^2}\Bigl( \tilde \Psi+\delta\Bigr)^2 \Biggl] .
\label{038}
\end{eqnarray}
The functions $f$ and $\tilde \Psi$ are assumed to
vanish on the  null cone $C_{a}$.
Therefore $\Psi =\delta $,   $\partial_R\Psi =\partial_R\delta $  and
 $\partial_t\Psi =\partial_t\delta$  on
$C_{a}$. In such a case the rate  of the energy
change  reads
\begin{eqnarray}
&&(\partial_0+\partial_{r^*})E_{a}= \nonumber\\
&&-2\pi (1-{2m\over R})\Bigl[ ((1-{2m\over R}) h^2_-
  +{ 2 \delta^2\over R^2} \Bigl] .
\label{039}
\end{eqnarray}
The  energy loss  is equal to a line integral  along     $\tilde \Gamma_a$,
\begin{eqnarray}
&& \delta E_a\equiv  E_{a}- E_{\infty }=
 \nonumber\\ &&
2\pi \int_{a}^{\infty  } dr
  \Biggl[ (1-{2m\over r}) h^2_-
  +{ 2 \delta^2\over r^2} \Biggl] .
\label{040}
\end{eqnarray}
The  derivation of (\ref{036}) requires the use of
estimates (\ref{21}) and  (\ref{033}).  The calculation of the $\delta $-
related part of the right hand side of (\ref{040}) is straightforward and
it yields
\begin{eqnarray}
&&  4\pi \int_{a}^{\infty  } dr   {  \delta^2\over r^2}  \le \nonumber \\
&& 4\pi \beta_a(b) \Bigl( {2m\over a}\Bigr)^2{1-{1\over
 (1+ \kappa  )^{2\epsilon }}\over \eta^2_a (1-2\epsilon )^2}\times
 \Biggl[ C^2_4 {2\epsilon^2\over
(3-2\epsilon )(1-\epsilon )}+  C_5^2 +2C_4C_5{1-2\epsilon \over 3-2\epsilon }
\Biggr] .
\label{041}
\end{eqnarray}
In order to bound the contribution coming from the backscattered
radiation amplitude $h_-$ one needs  the estimate (\ref{21}).
A straightforward calculation   shows  that
\begin{eqnarray}
&& 2 \pi \int_{a}^{\infty  } dr (1-{2m\over r}) h^2_-  \le \nonumber \\
&&\pi \beta_a(b)    \Bigl( {2m\over a}\Bigr)^2{1-{1\over
(1+ \kappa  )^{2\epsilon }}\over 2\eta^2_a (1-\epsilon )}\times \nonumber \\
&&  \Biggl[ {C^2_4\over 2}+ C_1C_4\Biggl(
y \ln {b\over R(b)}+{1\over 2(1-\epsilon )}(1-y)\Biggr) + \nonumber \\
&&C^2_1 \Biggl( {y\over 2}\ln^2{b\over R(b)}-{y\over 2(1-\epsilon)}\ln {R(b)\over a}
(1-2\ln {b\over R(b)})  +{1\over 4(1-\epsilon )^2}(1-y) \Biggr] ,
\label{042}
\end{eqnarray}
where $y\equiv (a/R(b))^{2-2\epsilon }$.
Neglecting the negative term with $\ln ...$  and using the bounds of
Appendix B on $b/R(b)$, one arrives at a bound that in conjuction with
(\ref{041}) proves  Theorem 3.

{\bf Remark.} The above estimate depends on the parameter $\epsilon $, that
should be chosen in such a way as to optimize the bound. The exact value
of the  optimal $\epsilon $ depends on $a$ and $ \kappa $, but the value
$\epsilon =1/8$ is proven to yield satisfactory estimates.

\section{Dependence of backscatter on   the frequency of waves }

The coefficients $C_4$ - $C_6$ appearing in Theorem 3  change with $ \kappa $ but
remain finite in the whole $(0, \infty )$
 range of possible values of $ \kappa =(b-a)/a$.
 In the case when the support of the
initial radiation is very narrow, i. e., $ \kappa
 << 1$,  then  the coefficient
${1-{1\over (1+ \kappa  )^{2\epsilon }}\over  \eta^2_a  }\sim  \kappa $.
 I such a case one obtains that
\begin{equation}
{\delta E_a\over E_a^b} \le C\Bigl( {2m\over a}\Bigr)^2\kappa ,
\label{043}
\end{equation}
where $C$ is a constant.
In the   limit $ \kappa \rightarrow 0$  the ratio
 ${\delta E_a\over E_a^b}$
becomes 0; the backscattering is negligible in the case of initial pulses
of electromagnetic energy that are very narrow. And conversely,
the bound becomes bigger with increase of the width of the radiation pulse.
The physical meaning of that can be deduced with the help
of the Fourier transform
theory. The {\it similarity theorem} (\cite{Bracewell})
states that compression of the support of a function  corresponds  to
expansion of the frequency scale.
  If a support of
initial data is made narrow,  then the wavelengths scale of the pulse extends in
the direction of short lengths. Therefore
the  message  behind the  obtained results must be
  that high frequency radiation
is essentially unhindered by the effect of backscattering and that long waves
can be backscattered.

This dependence of the backscattering  on the wave length has been in fact
observed in the numerical investigation of the propagation of pulses of
scalar massless fields \cite{Regucki}. In this case halving of the
length  has led to a similar decrease of the fraction of the
diffused energy.

In the case of a black hole or a neutron star the scale is set
essentially by the Schwarzschild radius $R_S=2m$; waves with lengths much
shorter than $R_S$ are  not backscattered,
while waves of lengths $\sim R_S$ can reveal quite a strong effect.
Moreover,  one can show that the  $(2m/R)^2$  dependence of the
effect implies  that  most of the energy diffusion occurs
in regions that are not very far (as compared to the Schwarzschild radius)
from the center.

In order to exemplify the above remarks,  consider the diffusion
effect   in following two  cases. Assume  the same location
    $a=4R_S$, of both radiative dipoles and

i)     $ \kappa =1/8$ (i. e., the fundamental wavelength $R_S$) for a pulse I;

ii)  $\kappa =1/128$
(i. e., the fundamental wavelength $R_S/8$) for the  pulse II.

In the calculation  that is reported below,   $\epsilon $
is chosen to be 1/8, in accordance with the remark ending the preceding
section.  Then in the case I one obtains $\delta E_a/E_a^b<0.37$, while
in the case II (of shorter waves) one gets  $\delta E_a/E_a^b<0.004$.

The evolution equation (\ref{16}) can be written in another form
 as
\begin{equation}
(\partial_0 + \partial_{r^*})\Bigl( (1-{2m\over R})h_+\Bigr)
= (1-{2m\over R})
\Biggl[ {2\over R^2}
\delta + {6mf\over R^4 }
\Bigr) \Biggr] .
\label{044}
\end{equation}
where
\begin{equation}
h_{+}(R,t) ={1\over 1-{2m \over R}}(-\partial_0+\partial_{r^*})
\Bigl( \delta +f\Bigr) .
\label{045}
\end{equation}
is the intensity of the outgoing part of the  radiation.
The inequality (\ref{011}) can be written as follows, applying Lemma 1 and
the remark following it,
\begin{equation}
{|f(R)| \over R^{3/2}}\le C \sqrt{ E_a^b}
\Biggl( 1-\Bigl( {a\over b}\Bigr)^{2\epsilon }\Biggr) ;
\label{046}
\end{equation}
here $C$ is some constant. The integration of  (\ref{044}) along a null
geodesic $\tilde \Gamma_{a}$ yields now
\begin{equation}
(1-{2m\over R})h_{+}(R(t),t) -
(1-{2m\over a})h_{+}(R(0),t=0)\propto   {1\over a^{3/2}}
\Biggl( 1-\Bigl( {a\over b}\Bigr)^{2\epsilon }\Biggr) ,
\label{047}
\end{equation}
where the proportionality constant  depends only on $\epsilon $, $2m/a$
and the initial energy $E_a^b$.
Fixing the  energy  $E_a^b$, one notices that in the
regime  $(b-a)/a<<1$ the right hand side of (\ref{047}) is essentially
 zero. Thus the product $(1-{2m\over R})h_{+}$ is   constant. In  this case
one clearly sees the manifestation of the redshift - the rescaling
of the amplitude $h_{+}$
\begin{eqnarray}
&&h_+(\infty )=\eta_{a}h_+(a, t=0).
\nonumber\\ &&
\end{eqnarray}
 See also a   discussion of that
fact in a massless scalar field theory\cite{MENOM}.

\section{ Distance dependence of energy diffusion and sharpness of
the estimates.}

The bound of Theorem 3 depends on the source location - it contains, among
other factors, a square of the factor $2m/a$. Thus the bounds in question
decrease with the increase of $a$.
The dependence on the distance  can actually be much stronger.
 In order to see this,
consider the   dipole radiation II of ii), described in the preceding section, but
located at $a=4m$ (instead of $a=8m$, as assumed formerly). One  obtains
that   now $\delta E_a/E_a^b\approx 0.77$, instead of the former bound
$0.001$.
Numerical results concerning the propagation of massless scalar fields also
show that the backscattered energy
decreases rapidly with the increasing distance  \cite{MENOM} .

It is of interest to establish how  accurate the final estimate is.
Most of the inequalities derived in this paper are sharp,
in the sense, that one can find  examples that saturate  them.
Thus, for instance, the two null-line  integrals of Section V are estimated
sharply (the inequalities saturate in Minkowski space-time).
Similarly results of Appendices A and B are also exact; again, the
inequalities become equalities in Minkowski space-time.
The energy estimate of Section III is not sharp; but
the "loss of sharpness", to say, can be less than 25 \% (see the final remark
in Sec. III). The main source of unsharpness is the omission of negative
terms in a bound on $\delta $ (Sec. VII) and in bounds of diffused energy in
 Sec. VIII; but that becomes
insignificant with the decrease of $ \kappa $,
i. e., when the width of the pulse
becomes small in comparison to the Schwarzschild radius.
On the other hand, the combination of two exact steps can be associated
with some loss in the accuracy.

Taking this into account, it is quite likely, that in the case of sources
characterized by $ \kappa <1$ the bound in question  gives an order of the
diffused energy.
On the other hand, if $ \kappa >>1$, then the bound of Theorem 3
 becomes very inaccurate, with
\begin{eqnarray}
&& {\delta E_a\over E_a^b}\propto C\Bigl( {2m\over a}\Bigr)^2,
\label{X}
\end{eqnarray}
where $C\approx 10^4$. As mentioned before, the main source of unsharpness
is the omission of negative terms in a bound on $\delta $ (Sec. VII) and in
 bounds of  Sec. VIII.  A more accurate treatment would significantly
improve the estimates in the long wave regime.

\section{Discussion}

The main result of this paper, Theorem 3,   states that the dipole energy
diffusion due to the  backscattering depends on the square of $2m/a$,
where $m$ is the mass of the gravitational source and $a$ is a
location of the pulse of radiation.
Sections  IX and X show  that  the high-frequency radiation
  essentially is   not backscattered, but that the low-frequency radiation
can   manifest a significant diffusion effect. The last statement  is best
described in terms of the dimensionless parameter $\tilde \kappa
  \equiv R_S/\lambda $,
where $\lambda $ is the fundamental  radiation length.
 If $\tilde  \kappa   >>1$ then the
backscattering is negligible, but if  $\tilde  \kappa  \approx 1$ then
it can be significant. The above results
demonstrate that  the effect becomes negligible at distances much
bigger than the Schwarzschild radius of a central mass. That rules out most
stars as objects that can  induce observable backscattering effects. For
a star of a solar type and $\lambda \sim R_S$, for instance,   the   ratio
${\delta E_a\over E_a^b}$  can be at most $10^{-20}$.
In the case of white dwarves and $\lambda \sim R_S$ the bound (\ref{X}) gives
${\delta E_a\over E_a^b}<10^{-8}$. For long-wave radiation the bounds
are bigger - the effect even looks as  marginally relevant, for white dwarves,
when ${\delta E_a\over E_a^b} \sim 10^{-2}$.   However a sharper
estimate would lower that by several orders.

On the other hand two astrophysical compact objects, neutron stars and
(most likely) black holes, are not excluded as objects of interest.

The backscattering would damp the total luminosity produced in accretion
disks that exist in vicinities of compact objects,
 but since the most efficient regions  of the
 disks  are located at a distance of (at least) several
Schwarzschild radii, the effect would be  probably weak.
More relevant can be "echoes" - aftermaths of violent flashy eruptions,
produced by a part of  radiation reflected from a close vicinity of a
horizon of a black hole.
Numerical calculations done in the massless scalar fields propagation
suggest that the amplitude of the reflected radiation can constitute up to
20 \% of the incident one, assuming that the length of the wave is
comparable to the Schwarzschild radius of a black hole.

The  results of this   section  can be in principle  generalized
into the case of higher order multipoles.  The key point would
consist in showing analogues of the energy estimates of Sec. III that would
bound the higher multipole moments. That should lead to a variant of Theorem
3 valid under reservations similar to those expressed earlier.

An analysis similar to that of the present paper can be repeated also
in the case of a weak gravitational radiation produced in disks rotating
around Schwarzschildean black holes. The  conclusions concerning the
fraction of the diffused energy can be similar.

{\bf Acknowledgements.}  This work has been suported in part  by
the KBN grant 2 PO3B 010 16. The author is grateful to Niall O' Murchadha
for    reading of the manuscript, many discussions and valuable comments.
Thanks are also due to Peter Aichelburg for a useful remark and to Krzysztof
Roszkowski for his help in editing of this paper.

\vfill \eject

\begin{center}
\begin{figure}[!htb]
\unitlength 0.1mm
\begin{center}
\begin{picture}(1100,600)
\put(100,100){\vector(0,1){500}}
\put(100,100){\vector(1,0){900}}
\put(350,100){\line(1,2){250}}
\put(550,100){\line(1,2){114}}
\multiput(550,500)(73,-110){4}{\line(2,-3){46}}
\put(75,570){$t$}
\put(340,60){$a$}
\put(350,100){\circle*{10}}
\put(525,60){$R_0 (r)$}
\put(550,100){\circle*{10}}
\put(795,60){$R_1$}
\put(815,100){\circle*{10}}
\put(885,60){$b$}
\put(890,100){\circle*{10}}
\put(550,500){\circle*{10}}
\put(570,485){$(R,t)$}
\put(672,335){$(r,\tau)$}
\end{picture}
\end{center}
\caption{Solid lines $\widetilde{\Gamma}_a, \widetilde{\Gamma}_{R_0 (r)}$
 are outgoing null curves. Broken line is $\Gamma_{R_1}$, an ingoing null
 curve.}
\end{figure}
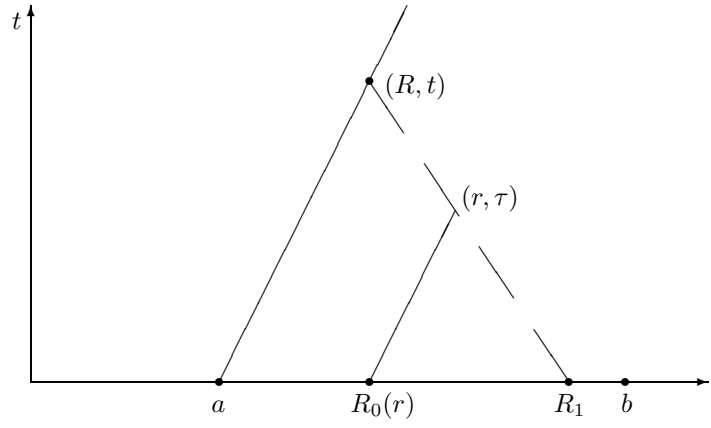
\end{center}
\begin{center}
\begin{figure}[!htb]
\unitlength 0.1mm
\begin{center}
\begin{picture}(1100,700)
\put(100,100){\vector(0,1){600}}
\put(100,100){\vector(1,0){900}}
\put(350,100){\line(1,2){300}}
\put(75,670){$t$}
\put(340,60){$a$}
\put(350,100){\circle*{10}}
\put(525,60){$R (b)$}
\multiput(550,100)(0,80){6}{\line(0,1){40}}
\multiput(550,500)(45,-60){7}{\line(3,-4){30}}
\multiput(600,600)(45,-60){7}{\line(3,-4){30}}
\put(600,600){\circle*{10}}
\put(900,200){\circle*{10}}
\put(910,180){$(R_1,s)$}
\put(620,600){$(R,t)$}
\put(845,60){$b$}
\put(850,100){\circle*{10}}
\put(850,100){\line(1,2){100}}
\end{picture}
\end{center}
\caption{Solid lines   $ \widetilde{\Gamma}_a, \widetilde{\Gamma}_b$
are outgoing null curves.
Broken lines are ingoing null curves  $\Gamma_b, \Gamma_{(R_1,s)}$.}
\end{figure}
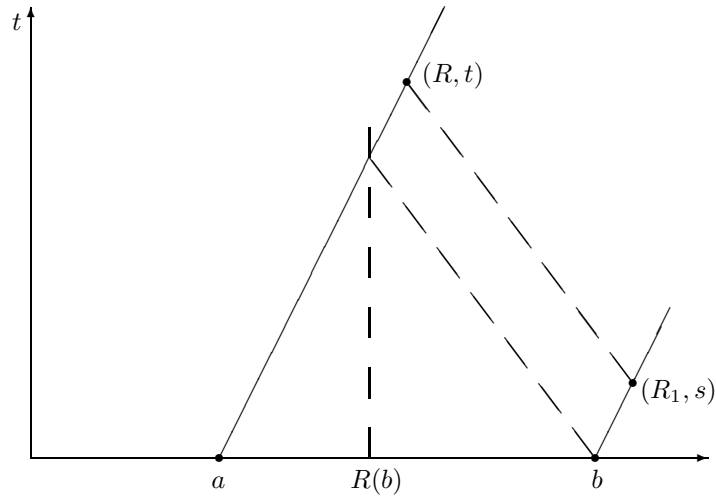
\end{center}

\vfill \eject

\section{Appendix A}

{\bf Lemma A.} Let $(R,t), R_1 \in \Gamma_{R_1 ,(R,t)}$,
 $R_1\ge R$ and $R>4m$.
 Then
\begin{equation}
{R\over R_1} \ge {a\over b}.
\end{equation}

{\bf Proof}.

 There are two separate cases that need to be considered.

i) If $R_1$ lies on the initial hypersurface then  $R_1\le b$ and $R\ge a$
and the inequality follows immediately.

ii) If  $(R_1,s)\in \tilde \Gamma_{b, (R_1,s)}$. In this case one has
\begin{equation}
2\Bigl( R_1 -R+2m\ln {R_1-2m\over R-2m}\Bigr) =
b-a +2m\ln {b-2m\over a-2m}
\label{A2}
\end{equation}
Define $X\equiv R_1-R$. One finds from (\ref{A2}) that
\begin{equation}
{d\over dR}\ln {X\over R} = {{2m\over R }\over 1-{2m\over R }}{1\over R+X}
-{1\over R}\le 0
\label{A3}
\end{equation}
provided that $R\le 4m$. Thus $X/R$ is a non-increasing function which
means that $R/R_1$ is a non-decreasing function and  $R/R_1 \ge R(b)/b$.
Since $R(b)\ge a$, one  arrives at the postulated inequality.

\section{Appendix B}
{\bf Lemma B}. Define $ \kappa \equiv (b-a)/a\le 0$.
Define $(R(b), t)$ as the the intersection point  of $\Gamma_b$ and
$\tilde \Gamma_a$. Then
\begin{equation}
{a+b\over 2}-m\kappa \le R(b)\le  {a+b\over 2}.
\end{equation}
{\bf Proof.} The relation (\ref{18c}) (see the main text), with $R_1=b$,
$r=R(b)$ and $R_0=a$, can be written  as                                       ,
\begin{equation}
a=2R(b)-b+ 2m\ln \Bigl( {(R(b)-2m)^2\over (a-2m)(b-2m)}
\Bigr) .
\label{A1}
\end{equation}
We will treat (\ref{A1}) as a relation between $b$ and $R(b)$, with fixed $a$.
Obviously $R(b)=b=a$ when $b=a$. One easily  finds that
\begin{equation}
\partial_bR(b)={1\over 2} {\eta_{_{R(b)}}\over \eta_b}.
\end{equation}
Notice that $R(b)\le b$. Thus $\partial_bR(b) \le 1/2$. On the other hand
$R(b)\ge a$. Thus   $\partial_bR(b) \ge (1/2)\eta_{_{R(b)}}\ge
(1/2)\eta_a$. The use of those two bounds on   $\partial_bR(b)$
and the initial condition $R(a)=a$ immediately imply
the Lemma.

\begin{references}
\bibitem{MENOMback} The condition for the smallness of backreaction
is of the type shown in \cite{MENOM} for the massless scalar field.
\end{references}
\end{document}